\documentclass[journal=cmatex,manuscript=article]{achemso}
\usepackage[version=3]{mhchem} % Formula subscripts using \ce{}

\usepackage[english]{babel}
\usepackage[utf8]{inputenc}
\usepackage{amsmath}
\usepackage{graphicx}
\usepackage[colorinlistoftodos]{todonotes}
\usepackage{caption}
\usepackage{subcaption}
\usepackage{amsfonts}
\usepackage{booktabs}
\usepackage{siunitx}
\usepackage{cleveref}
\usepackage{adjustbox}

\author{Yao Cai}
\email{yaocai@berkeley.edu}

\author{Wei Xie} 

\author{Hong Ding}
\affiliation{Department of Materials Science and Engineering, University of California, Berkeley, California, USA}

\author{Yan Chen}
\altaffiliation{School of Materials Science and Engineering, Nanyang Technological University, Singapore}

\author{Thirumal Krishnamoorthy}
\altaffiliation{School of Materials Science and Engineering, Nanyang Technological University, Singapore}

\author{Lydia H. Wong}
\altaffiliation{School of Materials Science and Engineering, Nanyang Technological University, Singapore}

\author{Nripan Mathews}
\altaffiliation{School of Materials Science and Engineering, Nanyang Technological University, Singapore}
\affiliation{Energy Research Institute @ NTU (ERI@N), Interdisciplinary Graduate School, Nanyang Technological University, Singapore}
\alsoaffiliation{Institute of Materials Research and Engineering (IMRE), Agency for Science, Technology and Research (A*STAR), Singapore}

\author{Subodh G. Mhaisalkar}
\affiliation{Energy Research Institute @ NTU (ERI@N), Interdisciplinary Graduate School, Nanyang Technological University, Singapore}
\altaffiliation{School of Materials Science and Engineering, Nanyang Technological University, Singapore}

\author{Matthew Sherburne}

\author{Mark Asta}
\affiliation{Department of Materials Science and Engineering, University of California, Berkeley, California, USA}
\email{mdasta@berkeley.edu}

\title{Computational Study of Halide Perovskite-Derived A$_2$BX$_6$ Inorganic Compounds: Chemical Trends in Electronic Structure and Structural Stability}

\begin{document}

\begin{tocentry}
\includegraphics{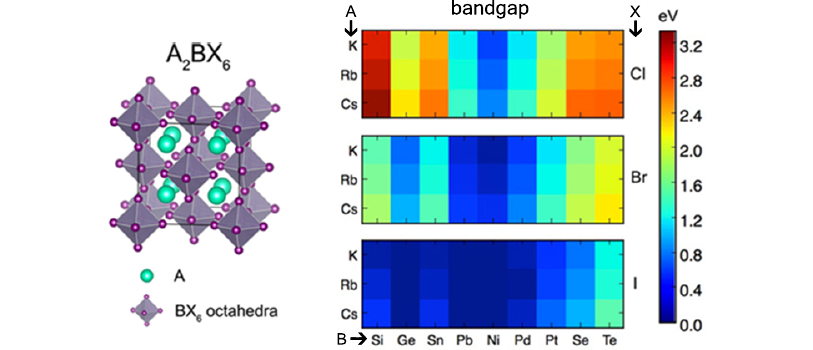}
\end{tocentry}

\begin{abstract}
The electronic structure and energetic stability of A$_2$BX$_6$ halide compounds with the cubic and tetragonal variants of the perovskite-derived K$_2$PtCl$_6$ prototype structure are investigated computationally within the frameworks of density-functional-theory (DFT) and hybrid (HSE06) functionals. The HSE06 calculations are undertaken for seven known A$_2$BX$_6$ compounds with A = K, Rb and Cs, and B = Sn, Pd, Pt, Te, and X = I. Trends in band gaps and energetic stability are identified, which are explored further employing DFT calculations over a larger range of chemistries, characterized by A = K, Rb, Cs, B = Si, Ge, Sn, Pb, Ni, Pd, Pt, Se and Te and X = Cl, Br, I. For the systems investigated in this work, the band gap increases from iodide to bromide to chloride. Further, variations in the A site cation influences the band gap as well as the preferred degree of tetragonal distortion. Smaller A site cations such as K and Rb favor tetragonal structural distortions, resulting in a slightly larger band gap. For variations in the B site in the (Ni, Pd, Pt) group and the (Se, Te) group, the band gap increases with increasing cation size. However, no observed chemical trend with respect to cation size for band gap was found for the (Si, Sn, Ge, Pb) group. The findings in this work provide guidelines for the design of halide A$_2$BX$_6$ compounds for potential photovoltaic applications.
\end{abstract}

\section{Introduction}
Since the initial discovery of lead halide perovskite compounds as solar absorbers in photovoltaic devices\cite{kojima_organometal_2009, lee_efficient_2012,kim_lead_2012}, the power conversion efficiencies (PCEs) achieved with these materials has increased steadily and currently reaches 22.1\%\cite{nrel_2017}. These high PCEs have motivated significant efforts aimed ultimately at the commercial application of lead-based halide perovskites for solar power conversion.  For such applications, two issues that continue to receive considerable attention are the toxicity of lead, which can be leached out of APbX$_3$ compounds due to their aqueous solubility \cite{Zhao:2015eo}, and the poor chemical stability in air \cite{Zhao:2015eo}.  The first of these two issues has motivated research into the use of alternative Sn/Ge-based perovskite compounds, although limited PCEs of 6\% or less have been demonstrated with these materials to date \cite{Noel:2014, Hao:2014cg, chakraborty_rational_2017,krishnamoorthy_lead-free_2015}. Further, like their Pb-based counterparts, Sn/Ge-based perovskite compounds also have been found to suffer from poor chemical stability \cite{Noel:2014, Hao:2014cg, krishnamoorthy_lead-free_2015}.

\begin{figure}[htb]
\centering
\includegraphics[width=0.6\textwidth]{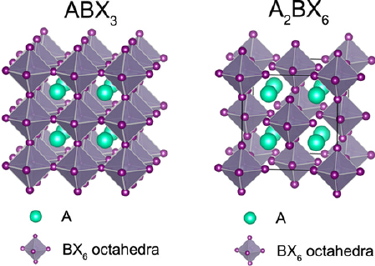}
\caption{\label{fig:structure}Crystal structures of $ABX_3$(left) and $A_2BX_6$(right).}
\end{figure}

The inorganic compound Cs$_2$SnI$_6$ has received recent attention as an alternative to Sn-based halide-perovskites for photovoltaic device applications.  In comparison with Sn and Pb based halide perovskites, Cs$_2$SnI$_6$ has been shown to feature enhanced stability in ambient environments\cite{Lee:2014kq, saparov_thin-film_2016,kaltzoglou_optical-vibrational_2016}, correlating with the presence of a more stable higher oxidation state for Sn in this compound (formally 4+ in Cs$_2$SnI$_6$ compared with 2+ in CsSnI$_3$). The crystal structure of Cs$_2$SnI$_6$ can be described as a defect variant of perovskite, with half of the Sn atoms removed, as illustrated in Fig.~\ref{fig:structure}.  In this figure, the SnX$_6$ octahedra in the Cs$_2$SnI$_6$ compound can be seen to be isolated, in contrast to the corner-sharing arrangement characterizing the perovskite structure. Associated with this structural change, the Sn-I bond lengths are shorter in the Cs$_2$SnI$_6$ compound, which has been correlated with its enhanced chemical stability \cite{Xiao:2015cua}. 

In addition to its desirable stability, the electronic and optical properties of Cs$_2$SnI$_6$ also have been demonstrated to be attractive for photovoltaic device applications. Specifically, Cs$_2$SnI$_6$ is a direct-gap semiconductor that displays strong visible light absorption \cite{saparov_thin-film_2016,qiu_unstable_2017}.  The measured band gaps reported range from 1.6 eV\cite{saparov_thin-film_2016} to 1.48 eV\cite{qiu_unstable_2017} to 1.3 eV\cite{Lee:2014kq}. The valence band and conduction band are well dispersed, with dominant characters based on I-$p$ and hybrid I-$p$/Sn-$s$ orbitals, respectively\cite{Lee:2014kq}. A high electron mobility of 310 cm$^2$/V$*$s and a sizable hole mobility 42 cm$^2$/V$*$s were reported in bulk compounds \cite{Lee:2014kq}. The Cs$_2$SnI$_6$ compound has been reported to display intrinsic $n$-type conductivity\cite{Lee:2014kq,saparov_thin-film_2016} (with carrier concentrations of 10$^{14}$ cm$^{-3}$ and 5$*$10$^{16}$ cm$^{-3}$), and it has been shown that it can be doped $p$-type with SnI$_2$\cite{Lee:2014kq} (with carrier concentrations of 10$^{14}$ cm$^{-3}$), demonstrating the ambipolar nature of this material \cite{Lee:2014kq}.  As an initial attempt, PCEs of about 1\% in air have been recently demonstrated in photovoltaic devices employing this material as a photoabsorber \cite{qiu_unstable_2017}.

As a means for further optimizing the properties of Cs$_2$SnI$_6$ for device applications, substitutional alloying is expected to be a useful strategy, as the A$_2$BX$_6$ structure can be readily doped with different impurity ions, especially in the 6-fold coordinated tetravalent cation site\cite{brik_modeling_2011}.   Studies investigating the effect of chemical substitutions have been limited to date, but they have clearly demonstrated the possibility of tuning electronic properties. For example, in a study comparing hole-transport properties in Cs$_2$SnI$_6$, Cs$_2$SnBr$_6$ and Cs$_2$SnCl$_6$ systematic variations in the optical and transport properties were observed to be correlated with the substitution of halogen anions having different size and electronegativity \cite{kaltzoglou_optical-vibrational_2016}. In another study the alloying of Cs$_2$SnI$_6$ with Cs$_2$TeI$_6$, to form Cs$_2$(Sn$_{1-x}$Te$_x$)I$_6$ solid solutions, was found to lead to reduced mobility and reduced defect tolerance \cite{maughan_defect_2016}. While these studies have not yet resulted in improved materials for device applications, they have clearly demonstrated the importance of the B-X chemistry in defining optical and transport properties.

To guide further studies aimed at the use of chemical substitutions to optimize properties for photovoltaic applications, in this work we employ density-functional-theory (DFT) based computational methods to explore trends underlying the variation of electronic structure and structural stability of A$_2$BX$_6$ compounds with chemical composition.  We begin by employing hybrid-functional (HSE06) methods to calculate and analyze the electronic structures of seven known inorganic compounds with varying A(=K, Rb, Cs) and B(=Pt, Pd, Sn, and Te) site cations, for the case of X=I.  We use these results as benchmarks to demonstrate the ability of computationally more efficient semi-local DFT functionals(Perdew-Burke-Ernzerhof parametrization of the generalized gradient approximation, or GGA-PBE) to capture chemical trends. Using the GGA-PBE method we extend the study to consider a wider range of chemistries on the A (A= K, Rb, Cs), B (B= Si, Ge, Sn, Pb, Ni, Pd, Pt, Se, Te) and X sites (X = Cl, Br, I). 

The results yield the following chemical trends. For variations in the halide anion from Cl to Br to I, the band gap and effective mass are found to decrease, while variations in the A site cation from Cs to Rb to K, lead to an increase in tetragonal distortion and an associated increase in band gap and effective mass. For variations in the B site in the group of Si, Ge, Sn, Pb, there is no observed chemical trend with respect to cation size for band gap, with larger values for Si and Sn and smaller values for Ge and Pb. For variations in the B site in the (Ni, Pd, Pt) group the band gap increases with increasing cation size. For variations in the B site in the group of Se and Te, the band gap increases with increasing size.

\section{Approach}
\subsection{Structures and Chemistries Considered}\label{ssec:approach1}

In this section we provide a brief description of the chemical compositions of the A$_2$BX$_6$ inorganic compounds considered in the present computational studies.  We focus initially on compounds with the halide anion X=I and consider seven compounds with varying alkali A-site cations and tetravalent B-site cations:  K$_2$PtI$_6$, Rb$_2$PtI$_6$, Cs$_2$PtI$_6$, Cs$_2$PdI$_6$, Rb$_2$SnI$_6$, Cs$_2$SnI$_6$ and Cs$_2$TeI$_6$.  Each of these compounds has been synthesized and characterized experimentally, with crystal structure parameters tabulated in the International Crystal Structure Database(ICSD)\cite{belsky_new_2002,bergerhoff_inorganic_1983}. All of these compounds have the cubic (space group Fm$\bar{3}$m) structure illustrated in Fig.~\ref{fig:structure} (b), with the exception of K$_2$PtI$_6$, which is tetragonally distorted (space group P4/mnc). From the DFT-GGA results available through the Materials Project \cite{Jain2013} these chemistries are expected to yield a range of band gap values that is relevant for photovoltaic applications. For each of these seven compounds we have undertaken computational studies of the electronic structure for both cubic and tetragonal polymorphs, employing the Heyd-Scuseria-Ernerhof HSE06 \cite{heyd_hybrid_2003,heyd_erratum:_2006} hybrid functional, as detailed below. The calculated hybrid-functional results are compared with available experimental data, and are used to demonstrate the ability of semi-local DFT calculations, based on the Perdew-Burke-Ernzerhof \cite{Perdew1996, Perdew1997} Generalized-Gradient Approximation (PBE-GGA), to capture the main chemical trends. With PBE-GGA DFT methods we expand the range of chemistries considered, presenting results in what follows for A$_2$BX$_6$ compounds with A=K, Rb, Cs, B=Si, Ge, Sn, Pb, Ni, Pd, Pt, Se and Te, X=Cl, Br, I. 

\begin{figure}[htbp]
\centering
\includegraphics[width=0.6\textwidth]{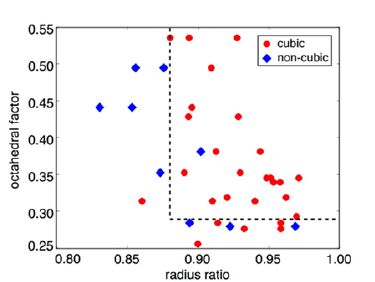}
\caption{\label{fig:structure_map} A structure map for known A$_2$B$X_6$ compounds with A = K, Rb, Cs, B = Si, Ge, Sn, Pb, Ni, Pd, Pt, Se, Te and X = Cl, Br, I. Crystal structures reported in the ICSD are indicated by the different symbols, with red circles and blue diamonds corresponding to cubic and non-cubic materials, respectively. For each material, the octahedral factor is calculated as $\frac{R_B}{R_X}$ and the radius ratio is calculated as $\frac{R_A}{D_{XX}-R_X}$, with $R_A$, $R_B$, $R_X$, $D_{XX}$ representing the radius of A site cation, radius of B site cation, radius of X site anion, and nearest neighbor X-X bond length, respectively. The X-X bond length is calculated from GGA-PBE relaxed structures for the cubic phase for each material. Shannon effective ionic radii for 12-coordinated A$^+$ cations, 6-coordinated B$^{4+}$ cations, and 6-coordinated X$^-$ anions\cite{shannon_revised_1976} are used for R$_A$, R$_B$ and R$_X$. The dashed line is a guide to the eye that separates the majority of the cubic perovskite structures and distorted non-cubic structures.}
\end{figure}

\begin{figure}[htbp]
\centering
\includegraphics[width=0.6\textwidth]{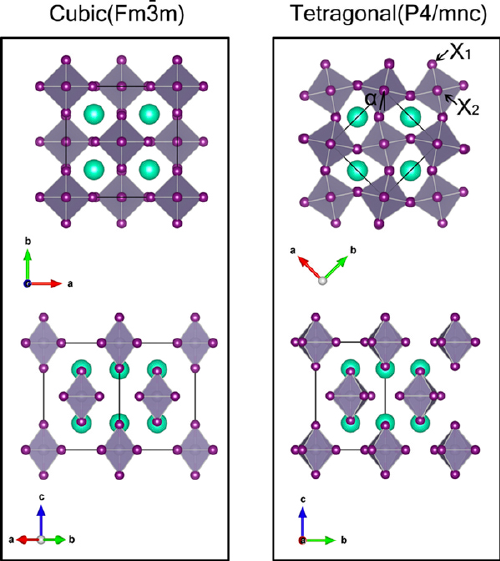}
\caption{\label{fig:cubic_tetra}Crystal structures for the A$_2$BX$_6$ compound in cubic Fm$\bar{3}$m (left) and tetragonal P4/mnc (right) polymorphs viewed from different directions. Lattice vectors are shown for each structure and orientation. The BX$_6$ octahedra are shaded, with the purple balls on the corners representing halide cations.  The green balls represent the A-site cations. In the tetragonal structure, octahedral rotation angle with respect to cubic structure is labeled as $\alpha$, equatorial halide as X$_1$, and apical halide as X$_2$.}
\end{figure}

To motivate the choice of crystal structures considered in this work, we show in Fig. \ref{fig:structure_map} a structure map similar to those used in studies of perovskite-based compounds\cite{li_can_2013}. The axes in this figure correspond to the octahedral factor and a radius ratio defined below. The octahedral factor is defined as the ratio between B cation radius and X anion radius. The radius ratio is the ratio between the radius of the A site cation and the size of the cavity formed by the neighboring halogen anions\cite{brown_crystal_1964}. For perovskite compounds the octahedral factor is used to empirically predict the formation of the BX$_6$ octahedron; and the tolerance factor is used to empirically predict the formation and distortion of the perovksite structure. Likewise, in the A$_2$BX$_6$ perovskite-derived structure, we can combine the octahedral factor and radius ratio to predict the formation and distortion of the structure. Small octahedral factors suggest that the formation of BX$_6$ octahedra are disfavored. A small radius ratio results in distortion of the cavity and a lower symmetry of the structure, or even totally different connectivity of the octahedra network. According to the survey of known A$_2$BX$_6$ compounds in the ICSD database\cite{belsky_new_2002,bergerhoff_inorganic_1983}, shown in Fig.\ref{fig:structure_map}, most known compounds are cubic (Fm$\bar{3}$m) and are indicated by red circles. Because of this, in this work, most of the trends are derived from considerations of cubic (Fm$\bar{3}$m) structures. On the other hand, as shown in Fig.\ref{fig:structure_map}, compounds with smaller radius ratio and smaller octahedral factor tend to form non-cubic structures, as indicated by blue diamond symbols. Among these non-cubic structure compounds, most adopt the tetragonal P4/mnc structure. In this work, the effect of tetragonal distortions and octahedral rotations are investigated for select compositions by comparing results for the cubic structure with those computed for the tetragonal P4/mnc structure. As shown in Fig. \ref{fig:cubic_tetra}, the two structures differ by the rotations of octahedra in the a-b plane.

\subsection{Computational Methods}
Calculations were carried out employing spin-polarized HSE06 and PBE-GGA based DFT methods using the Projector Augmented Wave (PAW) method \cite{Blochl1994}, as  implemented in the Vienna Ab initio simulation package (VASP)\cite{Kresse1996,kresse_efficiency_1996, kresee1993}. The PAW potentials used in the calculations are the same as those underlying the data provided in Materials Project\cite{Jain2013}, in order to facilitate comparisons of the results available through this database. The energy cutoff for the plane wave basis for all compounds was set to 520 eV. For cubic structures, self-consistent calculations were carried out with a gamma-centered k-point mesh of 6$\times$6$\times$6. Based on convergence tests for Cs$_2$SnI$_6$, Rb$_2$TeI$_6$ and K$_2$PtI$_6$, this choice of plane-wave cutoff and k-point density is found to be sufficient to provide total energies converged to within 1 meV/atom, lattice constants within 0.01 \AA, and band gaps within 1 meV. For tetragonal structures, self-consistent calculations were carried out with a gamma-centered k-point mesh of 4$\times$4$\times$4, to maintain a similar k-point density as used in cubic structures. For relaxation and density of state calculations, the tetrahedron method with Blöchl corrections was used for k-space integration. To check the importance of relativistic effects on band gaps, spin-orbit coupling is included using the standard approach in VASP for three representative compounds with heavier elements. The self-consistency iterations were performed until the energy was converged to within $1\times 10^{-5}$ eV. The structural relaxations were undertaken until the forces were converged within 0.01 eV/\AA  for the GGA-PBE calculations, and within 0.05 eV/\AA  for HSE06. For the compounds containing B-site transition-metal cations (B=Ni, Pd, and Pt) with unfilled $d$ shells in the 4+ charge state, we ran spin-polarized calculations considering both high-spin and low-spin configurations, finding the low-spin (zero local moment) states to be lowest in energy.

The average effective masses\cite{Hautier:1ea} were calculated using the BoltzTrap code\cite{Madsen:2006fj} and the pymatgen package\cite{ong_python_2013, Hautier:1ea}. Gamma-centered 20$\times$20$\times$20 and 18$\times$18$\times$12 k-point meshes were used for GGA-PBE band structure calculations of cubic and tetragonal structures, respectively. The band structures were then used as input to BoltzTrap code to calculate the conductivity tensor $\frac{\mathbf{\sigma}(T,\mu)}{\tau}$, with $\tau$ the constant relaxation time, T set to room temperature (300K), and $\mu$ the Fermi level. Then averaged effective masses were calculated as $\mathbf{m}=\frac{ne^2\tau}{\mathbf{\sigma}}$. As mentioned by Hautier et al\cite{Hautier:1ea}, the effective mass defined in this way is an average of $\mathbf{m}(i,\mathbf{k})$ around the Fermi level $\mu$, with i the index of band, $\mathbf{k}$ the wave vector. For electron effective mass, $\mu$ was set at the conduction band minimum; for hole effective mass, $\mu$ was set at the valence band maximum. Convergence of the effective mass values with respect to the density of k-points over which the band structure was sampled was carefully tested and the values were found to be converged to within 1\% using a 20$\times$20$\times$20 mesh.

\section{Results and discussion}

\subsection{Calculated Results for Experimentally Reported A$_2$BI$_6$ Compounds}

\begin{table}[htbp]
  \centering
  \begin{adjustbox}{max width=\textwidth}
    \begin{tabular}{rrrrr} 
    \toprule
    {Compound} & {Structure} &{GGA-PBE ($\AA$)} & {HSE06 ($\AA$)} & {Exp. ($\AA$)} \\\midrule
    {Cs$_2$SnI$_6$}&{cubic}  &a=8.53 &a=8.46 &a=8.235\cite{maughan_defect_2016}\\
    {Rb$_2$SnI$_6$}&{cubic}  &a=8.40 &a=8.32 &a=8.217\cite{Werker1939}\\
    %{Rb$_2$SnI$_6$}&{tetragonal}  &a=8.097 &a=8.025 &a=8.069\cite{} \\
    %{}        &{}       &c=12.621 &c=12.499 &c=11.786\cite{}      \\
    {Cs$_2$PdI$_6$}&{cubic}  &a=8.28 &a=8.21 &a=8.013\cite{schupp_crystal_2000}\\
    {Cs$_2$PtI$_6$}&{cubic}  &a=8.32 &a=8.26 &a=8.038\cite{Thiele:1983} \\
    {Rb$_2$PtI$_6$}&{cubic}  &a=8.16 &a=8.11 &a=7.932\cite{Thiele:1983}\\
    {K$_2$PtI$_6$}&{tetragonal}  &a=7.73  &a=7.69  &a=7.717\cite{Thiele:1983}\\
    {}        &{}      &c=12.15 &c=12.05 &c=11.454\cite{Thiele:1983}\\
    {Cs$_2$TeI$_6$}&{cubic}  &a=8.60 &a=8.51 &a=8.275\cite{maughan_defect_2016} \\\bottomrule
    \end{tabular}%
    \end{adjustbox}%
  \caption{\label{tab:lattice}Calculated and experimentally measured lattice parameters of seven A$_2$BI$_6$ compounds reported in the ICSD. The second and third columns list calculated results obtained by semi-local GGA-PBE and hybrid HSE06 functionals, respectively.  The fourth column lists experimental values at room temperature. The structures labeled in the second column correspond to the cubic (Fm$\bar{3}$m) and tetragonal (P4/mnc) polymorphs described in the text.}
\end{table}%

Table \ref{tab:lattice} lists calculated results for the lattice constants of the seven reported A$_2$BI$_6$ compounds, along with experimentally measured values at room temperature. The calculated and measured results agree reasonably well with the GGA-PBE values larger than measurements by as much as 3.3\%, those from HSE06 being in slightly better agreement with deviations of up to 2.5\%. Table \ref{tab:bonds} shows that no obvious difference of B-X bond length is found between cubic and tetragonal structures. The last column of Table \ref{tab:bonds} lists the octahedral rotation angle $\alpha$ as illustrated in Fig. \ref{fig:cubic_tetra}. Among the seven compounds, K$_2$PtI$_6$ has the largest $\alpha$, consistent with the fact that it is the only compound that is experimentally observed to form in the tetragonal structure at room temperature.

\begin{table}[htbp]
  \centering
  \begin{adjustbox}{max width=\textwidth}
    \begin{tabular}{rrrrr} \toprule
     {} & {cubic}  & {}  & {tetragonal} & {} \\
    {Compound}  & {$B-X$/\AA}  &  {$B-X_1$/\AA} & {$B-X_2$/\AA} & $\alpha/\deg$  \\\midrule
    {Cs$_2$SnI$_6$}  & 2.87 & 2.87 & 2.88 & 5.1\\
    {Rb$_2$SnI$_6$}  & 2.86 & 2.86 & 2.89 & 9.4\\
    {Cs$_2$PdI$_6$}  & 2.69 & 2.69 & 2.69 & 0.3\\
    {Cs$_2$PtI$_6$}  & 2.69 & 2.69 & 2.69 & 0.3\\
    {Rb$_2$PtI$_6$}  & 2.68 & 2.69 & 2.68 & 8.7\\
    {K$_2$PtI$_6$}  & 2.68 & 2.69 & 2.68 & 12.0 \\
    {Cs$_2$TeI$_6$}  & 2.93 & 2.94 & 2.94 & 3.1\\\bottomrule
    \end{tabular}%
    \end{adjustbox}%
  \caption{\label{tab:bonds}Calculated bond lengths and rotation angles of seven A$_2$BI$_6$ compounds for both cubic (Fm$\bar{3}$m) and tetragonal (P4/mnc) polymorphs from HSE06. The second column lists the B-X bond length for the cubic structures. The last three columns list equatorial B-X bond length, apical B-X bond length and octahedral rotation angle $\alpha$ for tetragonal structures, respectively. The equatorial atom X$_1$, apical atom X$_2$ and $\alpha$ are illustrated in Fig.\ref{fig:cubic_tetra}. }
\end{table}%

\begin{table}[htbp]
  \centering
  \begin{adjustbox}{max width=\textwidth}
    \begin{tabular}{rllll} \toprule
    {Compound}  &  {$E_{g(HSE-cubic)}/eV$}  &  {$E_{g(HSE-SOC-cubic)}/eV$} & {$E_{g(HSE-tetragonal)}/eV$} & {$E_{g(exp)}/eV$} \\\midrule
    {Cs$_2$SnI$_6$}  & \bf{1.172} $\left(\Gamma-\Gamma\right)$  &\bf{1.011}$\left(\Gamma-\Gamma\right)$ &  1.216$\left(\Gamma-\Gamma\right)$  & 1.3 - 1.6 \cite{saparov_thin-film_2016,qiu_unstable_2017,Lee:2014kq}\\
    {Rb$_2$SnI$_6$}  & \bf{1.021} $\left(\Gamma-\Gamma\right)$  & & 1.510$\left(\Gamma-\Gamma\right)$   \\
   {Cs$_2$PdI$_6$}  &  \bf{0.858} $\left(\Gamma-X\right)$ & & 0.850$\left(\Gamma-M\right)$   \\
    {Cs$_2$PtI$_6$}  &  \bf{1.472} $\left(\Gamma-X\right)$ &\bf{1.340}$\left(\Gamma-X\right)$& 1.463$\left(\Gamma-\Gamma\right)$ \\
    {Rb$_2$PtI$_6$}  & \bf{1.303} $\left(\Gamma-X\right)$  & & 1.574$\left(\Gamma-M\right)$  \\
    {K$_2$PtI$_6$}  & 1.195 $\left(\Gamma-X\right)$  & & \bf{1.649}$\left(\Gamma-M\right)$  \\
    {Cs$_2$TeI$_6$}  &  \bf{2.187} $\left(X-L\right)$ &\bf{1.976}$\left(X-L\right)$& 2.080$\left(A-R\right)$ &1.5 - 1.59 \cite{maughan_defect_2016, peresh_preparation_2002}  \\\bottomrule
    \end{tabular}%
    \end{adjustbox}%
  \caption{\label{tab:gap}Band gaps calculated by the HSE06 functional for seven A$_2$BI$_6$ compounds reported in the ICSD. The second and third columns give calculated results for cubic structures with and without spin-orbit-coupling (SOC) contributions includes, respectively.  The fourth column gives results for the tetragonal structure.}
\end{table}%

Band gaps and effective masses of both cubic and tetragonal structures calculated by HSE06 are listed in Tables \ref{tab:gap} and \ref{tab:mass}, respectively. In these tables, results for the structures that are reported to be stable at room temperature are indicated in bold font. The HSE06 calculated band gaps listed in Table \ref{tab:gap} span the range of 0.8 to 2.2 eV. For Cs$_2$SnI$_6$, Cs$_2$PtI$_6$ and Cs$_2$TeI$_6$, the effects of spin-orbit coupling (SOC) on the calculated bandgaps were computed and found to lead to at most a reduction of 0.2 eV.  For the other compounds, involving lighter ions, these effects are expected to be smaller. The calculated band gap of Cs$_2$SnI$_6$ including spin-orbit coupling is 1.011 eV, which is consistent with the value of 0.97 eV reported in the literature using the same computational approach\cite{maughan_defect_2016}. As pointed out by Maughan et al\cite{maughan_defect_2016}, the calculated band gap is the fundamental band gap, across which the transition is dipole forbidden in Cs$_2$SnI$_6$. This explains why measured band gap is significant larger than calculated values. The calculated band gap value for Cs$_2$TeI$_6$ is 0.4 eV larger than measured values, which is acceptable given the 0.3 eV mean absolute error of band gaps reported for HSE06 for a semiconductor test set\cite{heyd_energy_2005}. A HSE06+SOC calculation reported\cite{maughan_defect_2016} for Cs$_2$TeI$_6$ gives an indirect band gap of 1.83 eV. The 0.15 eV discrepancy with our calculated results listed in Table \ref{tab:gap} may reflect the smaller cutoff energy and k-point density used in the previous calculations.

The present calculated results display clear trends with the size of the A-site cation. Considering first the trends for the cubic structures, the band gap values decrease systematically as the A site varies from Cs to Rb to K. By contrast, the trend is the opposite in the tegragonal structure. For a given chemistry, the band gap increases in going from the cubic to the tetragonal structure, and the magnitude of this increase is larger for the compounds with the smaller A-site cation.  These trends will be discussed further below based on the bonding in these compounds.

\begin{figure}[htbp]
\centering
\includegraphics[width=0.6\textwidth]{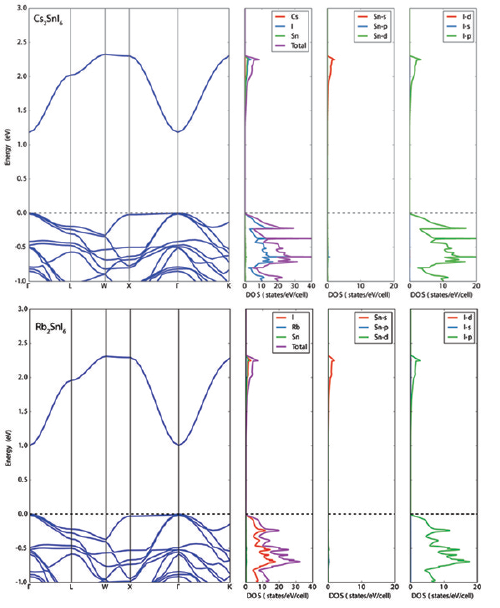}
\caption{\label{fig:Sn-dos} Band structures, total and projected densities of states, calculated by HSE06 for the experimentally observed A$_2$BX$_6$ compounds with B=Sn and X=I.  Results are plotted for Cs$_2$SnI$_6$ and Rb$_2$SnI$_6$ with the stable cubic (Fm$\bar{3}$m) structure.}
\end{figure}

\begin{figure}[htbp]
\centering
\includegraphics[width=0.6\textwidth]{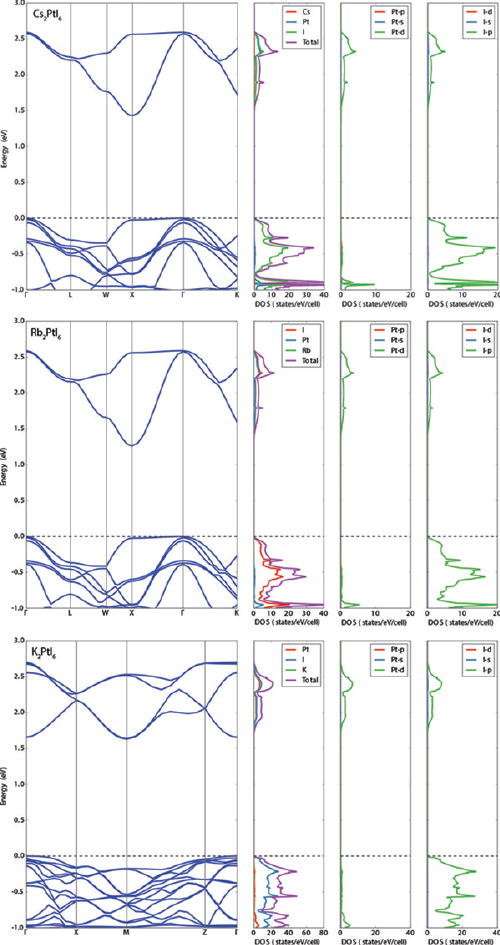}
\caption{\label{fig:Pt-dos} Band structures, total and projected densities of states, calculated by HSE06 for the experimentally observed A$_2$BX$_6$ compounds with B=Pt and X=I.  Results are plotted for Cs$_2$PtI$_6$ and  Rb$_2$PtI$_6$ in the stable cubic (Fm$\bar{3}$m) structure, and for K$_2$PtI$_6$ in the stable tetragonal (P4/mnc) structure.}
\end{figure}

\begin{figure}[htbp]
\centering
\includegraphics[width=0.6\textwidth]{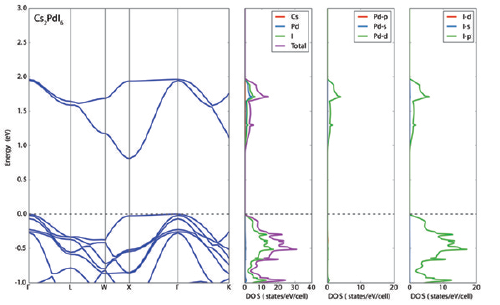}
\caption{\label{fig:Pd-dos} Band structures, total and projected densities of states, calculated by HSE06 for the experimentally observed Cs$_2$PdI$_6$ compound with the stable cubic (Fm$\bar{3}$m) structure.}
\end{figure}

\begin{figure}[htbp]
\centering
\includegraphics[width=0.6\textwidth]{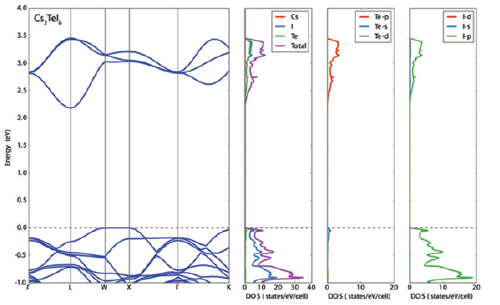}
\caption{\label{fig:Te-dos} Band structures, total and projected densities of states, calculated by HSE06 for the experimentally observed Cs$_2$TeI$_6$ compound with the stable cubic (Fm$\bar{3}$m) structure. }
\end{figure}

%discussion of each compound in detail
From Fig.\ref{fig:Sn-dos} to Fig.\ref{fig:Te-dos} we present calculated HSE06 bandstructures all obtained neglecting SOC effects for each of the seven compounds listed in Tables \ref{tab:gap}. Each separate figure gives bandstructures for a set of compounds with a fixed B-site cation. Beginning with Sn-based compounds, Cs$_2$SnI$_6$ and Rb$_2$SnI$_6$ have both been synthesized and their structures characterized experimentally\cite{Werker1939}.   
Considering first the electronic structures of Cs$_2$SnI$_6$ and Rb$_2$SnI$_6$ in Fig. \ref{fig:Sn-dos}, both compounds are calculated to be direct gap semiconductors with valence bands derived primarily from I $p$ states, and conduction bands derived from hybridized Sn $s$ and I $p$ states. The direct nature of band gap in Cs$_2$SnI$_6$ is consistent both with previous ab initio calculations\cite{Lee:2014kq,xiao_ligand-hole_2015} as well as experiments\cite{Lee:2014kq,saparov_thin-film_2016}.

%According to the spin-polarized calculations, the ground states for these d6 compounds are nonmagnetic low spin states. 
We consider next the B=Pd and Pt systems Cs$_2$PdI$_6$, Cs$_2$PtI$_6$, Rb$_2$PtI$_6$, and K$_2$PtI$_6$.  The latter compound forms in the tetragonal structure while all of the others are cubic\cite{schupp_crystal_2000,Thiele:1983}. The calculated bandstructures in Fig. \ref{fig:Pt-dos} and Fig. \ref{fig:Pd-dos} show that these compounds are indirect gap semiconductors with valence bands derived primarily from I $p$ states, and conduction bands derived from hybridized Pd/Pt $d$ and I $p$ states. Note that their direct gaps at X points are only slightly larger than the fundamental gaps.

We consider finally the Cs$_2$TeI$_6$ compound, which is experimentally observed to form in the cubic structure\cite{maughan_defect_2016}, with measured bandgaps reported ranging from 1.5 eV\cite{peresh_preparation_2002} to 1.59 eV\cite{maughan_defect_2016}. The measured bandgap values are smaller than the HSE06 calculated values(including SOC) by $\sim$ 0.4 eV. The calculated bandstructure in Fig. \ref{fig:Te-dos} shows that this compound is an indirect gap semiconductor with valence bands derived primarily from I $p$ states, and conduction bands derived from hybridized Te $p$ and I $p$ states. These results are consistent with the results reported by Maughan\cite{maughan_defect_2016}.

%effective mass discussion
% (from 2 papers including effective mass screening, one is using 1.5 and the other 1; as a reference for Si $mass_e=0.26, mass_h=0.39$, for $MAPbI_3$ $mass_e=0.19, mass_h=0.25$). 
We consider next the results for effective masses listed in Table \ref{tab:mass}. In all seven compounds, holes are found to be heavier than electrons. This result can be understood based on the fact that the valence bands are derived from unhybridized halogen $p$ orbitals, which are less dispersed compared with the conduction bands that are derived from anti-bonding states between halogen anions and B site cations. The trend obtained here is contrary to the case in halide perovskites, where holes are lighter than electrons\cite{huang_electronic_2013}. The trends in the effective mass values listed in Table \ref{tab:mass} are correlated with those for the band gaps.  Specifically, in cubic structures, for a given B-site cation, reducing the size of the A-site cation leads to a decrease in both the effective mass and the band gap, consistent with the expectations of $k\cdot p$ theory\cite{yu_cardona}. In tetragonal structures, for a given B-site cation, reducing the size of the A-site cation leads to an increase in both electron effective mass and the band gap. The variations of hole effective masses are more complex because of the presence of multiple valence bands at the band edge.

\begin{table}[htbp]
  \centering
  \begin{adjustbox}{max width=\textwidth}
    \begin{tabular}{rrrrrrr} \toprule
    {Compound}  & {$m_{e(cubic)}^*$}  &  {$m_{h(cubic)}^*$} &  {$m_{e(tetra)[100]}^*$}  &{$m_{e(tetra)[001]}^*$} & {$m_{h(tetra)[100]}^*$} & {$m_{h(tetra)[001]}^*$}  \\\midrule
    {Cs$_2$SnI$_6$}  & \bf{0.33} & \bf{1.50} & 0.44 & 0.42&1.27 & 2.79\\
    {Rb$_2$SnI$_6$}  & \bf{0.29}  & \bf{1.34} & 0.65 & 0.64 & 2.61 & 2.17\\
    {Cs$_2$PdI$_6$}  &  \bf{0.47} & \bf{1.37} & 0.46 &0.47 & 1.32& 1.50\\
    {Cs$_2$PtI$_6$}  &  \bf{0.51} & \bf{1.45} & 0.52 &0.53 & 1.38& 1.57\\
    {Rb$_2$PtI$_6$}  &  \bf{0.45} & \bf{1.25} & 0.59 & 0.66 & 2.50&1.32\\
    {K$_2$PtI$_6$}  & 0.40  & 1.15 & \bf{0.84} & \bf{0.83}&\bf{1.79} & \bf{2.55}\\
    {Cs$_2$TeI$_6$}  &  \bf{0.40} & \bf{1.51} &0.39  &0.39 & 1.48& 1.55\\\bottomrule
    \end{tabular}%
    \end{adjustbox}%
  \caption{\label{tab:mass}Calculatd effective masses for seven A$_2$BI$_6$ compounds reported in the ICSD. The results for electrons and holes in the cubic structures are given in columns two and three, respectively. The remaining columns list results for electrons and holes in the tetragonal structure along the two independent crystallographic directions:  [100] and [001].}
\end{table}%

\subsection{Trends in band gaps and stability across broader compositional ranges}

\begin{figure}[htbp]
\centering
\includegraphics[width=0.6\textwidth]{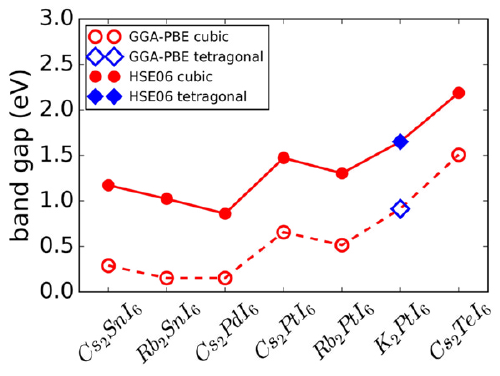}
\caption{\label{fig:pbe-hse-gap} Comparison between calculated band gaps obtained with the GGA-PBE and HSE06 methods. The HSE06 calculated band gaps for cubic and tetragonal structures are indicated by filled red circles and filled blue diamonds, respectively. The GGA-PBE calulated band gaps for cubic and tetragonal structures are indicated by open red circles and open blue diamonds, respectively. The solid and dashed lines are guides to the eye. }
\end{figure}

The HSE06 results in the previous section display clear trends in the electronic structure as the compositions of A$_2$BX$_6$ are varied with X=I.  In this section we investigate these trends over a broader range of 81 total compositions, considering A=(K, Rb, Cs), B=(Si, Ge, Sn, Pb, Ni, Pd, Pt, Se, Te), and X=(Cl, Br, I), using the computationally efficient GGA-PBE method. As shown in Fig. \ref{fig:pbe-hse-gap}, which compares results for band gaps obtained from GGA-PBE and HSE06 for the compounds considered in the previous section, the former reproduces the trends from the latter method quite well, even though the semi-local GGA-PBE functional systematically underestimates the band-gap values as expected.  The emphasis in the presentation of results in this section is specifically on compositional trends rather than absolute values for the bandgaps. In addition, we present results related to the relative structural stability of the cubic and tetragonal phases of the compounds.

\begin{figure}[htbp]
\centering
\includegraphics[width=0.8\textwidth]{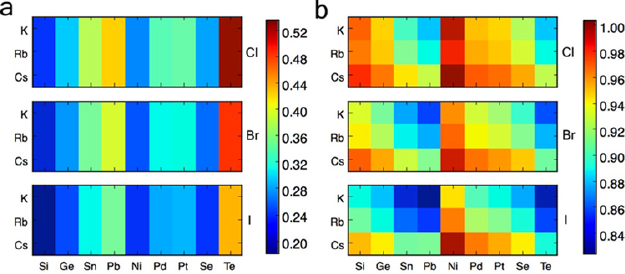}
\caption{\label{fig:structure_colorful} A color map of (a)octahedron factor and (b)radius ratio for 81 A$_2$BX$_6$ compounds with A = K, Rb, Cs, B = Si, Ge, Sn, Pb, Ni, Pd, Pt, Se, Te and X = Cl, Br, I. }
\end{figure}

\begin{figure}[htbp]
\centering
\includegraphics[width=0.5\textwidth]{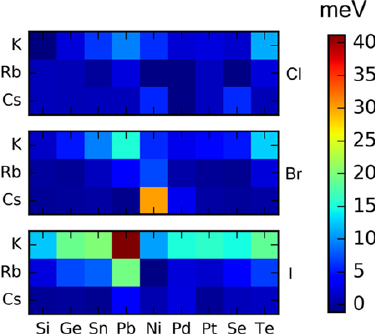}
\caption{\label{fig:energy_diff} A color map of the energy difference per atom of cubic minus tetragonal structures for 81 A$_2$BX$_6$ compounds with A = K, Rb, Cs, B = Si, Ge, Sn, Pb, Ni, Pd, Pt, Se, Te and X = Cl, Br, I. }
\end{figure}

We consider first the geometrical factors of octahedral factor and radius ratio introduced above. Figure \ref{fig:structure_colorful} plots these values for each of the 81 compositions considered, and the results provide insights into why only a small subset of these compositions have been observed to form A$_2$BX$_6$ compounds experimentally.  For example, it can been seen from Fig. \ref{fig:structure_colorful}(a) that B=(Si, Ge, Ni and Se) cations lead to small octahedral factors.  For these B-site cations only K$_2$SeBr$_6$, Rb$_2$SeCl$_6$, Cs$_2$SeCl$_6$ and Cs$_2$GeCl$_6$ have been reported to form experimentally according to the ICSD \cite{laubengayer_chlorogermanic_1940, engel_kristallstrukturen_2015,abriel_phase_1990}; no compounds with X=I have been reported, consistent with the smaller octahedral factors characterizing these compounds.  The results in Fig. \ref{fig:structure_colorful}(b) show that the radius ratio decreases systematically for A cations changing from Cs to Rb to K, and for X site cations varying from Cl to Br to I.  These results suggest that compounds with A=K and/or X=I are expected to display distorted (non-cubic) phases, which is consistent with the reported non-cubic crystal structures for K$_2$PtI$_6$, Rb$_2$TeI$_6$, K$_2$TeI$_6$ and K$_2$TeBr$_6$ in the ICSD \cite{Thiele:1983,abriel_crystal_1982,syoyama_x-ray_1972,brown_crystal_1964}. This argument is also supported by calculating the energy difference between cubic and tetragonal structures for the 81 compounds, as displayed in Fig. \ref{fig:energy_diff}. The energy of tetragonal K$_2$BI$_6$ is significantly lower than the energy of cubic K$_2$BI$_6$, indicating the greater tendency to form tetragonal structures for K$_2$BI$_6$ compounds.

%the trend of band gaps and bonding explanation
\begin{figure}[htbp]
\centering
\includegraphics[width=0.5\textwidth]{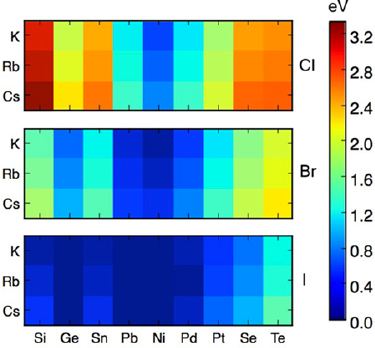}
\caption{\label{fig:bandgap}A color map of PBE-GGA calculated band gaps for 81 A$_2$BX$_6$ compounds in the cubic (Fm$\bar{3}$m) structure, with A = K, Rb, Cs, B = Si, Ge, Sn, Pb, Ni, Pd, Pt, Se, Te and X = Cl, Br, I. The three large blocks respond to Cl, Br and I compounds, respectively.}
\end{figure}

\begin{figure}[htbp]
\centering
\includegraphics[width=1.0\textwidth]{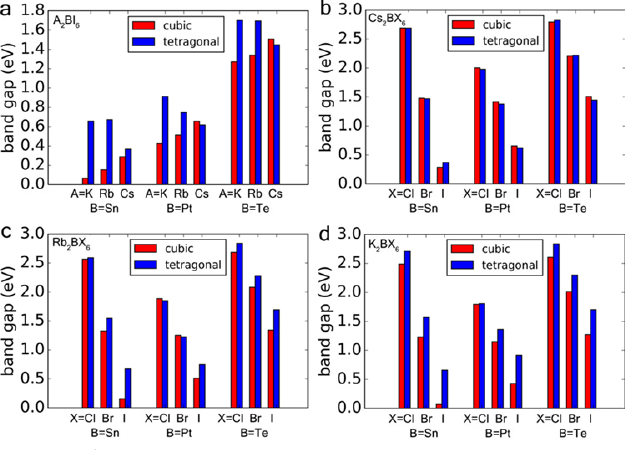}
\caption{\label{fig:trend}Chemical trends for GGA-PBE calculated band gaps in $A_2BX_6$(A=K, Rb, Cs, B=Sn, Pt, Te, X=Cl, Br, I) compounds with cubic (Fm$\bar{3}$m) and tetragonal (P4/mnc) structures. Band gap values for cubic and tetragonal structures are indicated by red bars and blue bars, respectively. (a)band gap trends with varying size of A site cations in $A_2BI_6$(A=K, Rb, Cs, B=Sn, Pt, Te); (b)band gap trends with varying size of X site anions in $Cs_2BX_6$(B=Sn, Pt, Te, X=Cl, Br, I); (c)band gap trends with varying size of X site anions in $Rb_2BX_6$(B=Sn, Pt, Te, X=Cl, Br, I);  (d)band gap trends with varying size of X site anions in $K_2BX_6$(B=Sn, Pt, Te, X=Cl, Br, I).}
\end{figure}

Calculated band gaps obtained with the GGA-PBE functional for the 81 compounds in ideal cubic structures are plotted in Fig.\ref{fig:bandgap}. The compounds on the 'blue' side have smaller band gaps.  Based on a comparison of these GGA-PBE results with the more accurate HSE06 calculations presented in the previous section these materials may have band gaps in the right range for sunlight absorber calculations. Similarly, the compounds such as the chloride shown in red in Fig. \ref{fig:bandgap} are expected to be large gap semiconductors/insulators. General chemical trends are apparent in the calculated band gaps with increasing size of A site cation and halogen anion, as demonstrated more explicitly in Fig. \ref{fig:trend}, which includes results for both cubic and tetragonal structures. To reduce the complexity of this figure, only Sn, Pt and Te compounds are shown, however the trends in Fig. \ref{fig:bandgap} for compounds with other B-site cations are similar.

The calculated band gaps are shown in Fig. \ref{fig:trend} (b)-(d) to decrease with increasing size of the halide anion, i.e., from Cl to Br to I.  This trend occurs for both cubic and tetragonal structures, and across all of the B and A site combinations presented.  This trend can be understood based on the electronic states at the conduction and valence bands. Taking B=Sn compounds as an example, the conduction band is composed of anti-bonding states between Sn $s$ orbitals and halide $p$ orbitals.  With increasing size of the halide anion the Sn-X bonds increase systematically, correlating with a decrease in the splitting between bonding and anti-bonding splitting and a concomitant lowering of the CBM.  The VBM is composed of states that are primarily derived from the halide anion $p$ orbitals, and these states are expected to rise in energy with the decreasing electronegativity going from Cl to Br to I.  The trend in decreasing band gap with increase size of the halide anion can thus be understood as resulting from a combination of a lower CBM and higher VBM. The bonding picture presented above is consistent with that discussed by Xiao et al\cite{xiao_ligand-hole_2015}.

The effects of variations of the A-site cation on the calculated band gaps are illustrated in Fig. 11 (a) for the case of X=I.  The effects are seen to be weaker than those resulting from variations in the halide anion, and they are shown to be qualitatively different for cubic and tetragonal phases.  Specifically, the calculated band gaps are found to increase and decrease for cubic and tetragonal structures, respectively, with increasing A-site cation size.  For the cubic phases the results can be understood as arising from the effect of the A-site cation on the distance between neighboring I sites.  As the size of the A-site cation increases, the distance between neighboring I sites increases, while the B-X bond lengths in the BI$_6$ octahedra remain largely unchanged.  Increasing I neighbor distances correlate with a narrowing of the I-p band and thus a lowering of the VBM, consistent with the increase in band gap from K to Rb to Cs.  

We consider next the trends with A-site cation for the tetragonal structures.  For the smaller A-site cations (K and Rb) the tetragonal structure is lower in energy and the band gap larger than that for the corresponding cubic phase.  With decreasing size of the A site cation, the degree of octahedral rotation (c.f., Fig. 2) increases, which leads to a decrease in the bonding strength between neighboring halide-ions as their $p$ orbitals increasingly point away from each other.  The larger band gap for the smaller A-site cation in the tetragonal phase thus correlates with a resulting narrowing of the valence band.

\section{Summary}
First-principles calculations employing the hybrid HSE06 method have been undertaken to compute the electronic structures of seven perovskite-derived A$_2$BI$_6$ compounds, considering B=Sn, Pd, Pt and Te cations. Calculated band gaps and effective masses of cubic structures decrease as the A site cation size decreases. These trends were explored over a broader range of A$_2$BX$_6$ halide chemistries, considering in total 81 combinations of A(=K, Rb, Cs), B=(Si, Ge, Sn, Pb, Ni, Pd, Pt, Se, Te) and X(=Cl, Br and I) ions, employing semi-local GGA-PBE calculations.  The results show that the trend of increasing band gaps with decreasing size of the halide anion holds across the compounds, for both cubic and tetragonal structures.  The effect of A-site cations is more complex.  Within the cubic structure, decreasing size of the A-site cation leads to a decrease in the calculated band gap, while also favoring structural distortion associated with the rotation of the BX$_6$ octahedra in the tetragonal phase, which has the effect of increasing the calculated band gap.

The trends identified in this computational study provide guidelines for the use of substitutional alloying as a means of tuning band gaps and structural stability for use of A$_2$BX$_6$ compounds in applications such as solar photo-absorbers.  For example, alloying of Cs$_2$BX$_6$ with Rb$_2$BX$_6$ could be expected to give rise to increasing rotation of the BX$_6$ octahedra and an increase in the band gap.  Similarly, as already demonstrated by McMeekin et al\cite{McMeekin151}, alloying of both A and X sites can be expected to be effective in tuning both bandgap and structural stability.  We note that alloying of B site Sn and Te cations has also been explored in this context \cite{maughan_defect_2016}.  The general trends in band gaps and structural stability identified in this computational study are anticipated to be helpful in guiding further work in these directions.

\begin{acknowledgement}
This work was funded by National Research Foundation (NRF), Singapore (CRP NRF2014NRF-CRP002-036) and the Singapore-Berkeley Research Initiative for Sustainable Energy (SinBeRISE) CREATE programme. This work made use of computational resources provided under the Extreme Science and Engineering Discovery Environment (XSEDE),  which is supported by the National Science Foundation  grant No. OCI-1053575.
\end{acknowledgement}

\bibliography{bib.bib}

\end{document}